\documentclass{article}
\usepackage{graphicx}
\usepackage[english]{babel}

\title{
\includegraphics[width=0.35\textwidth]{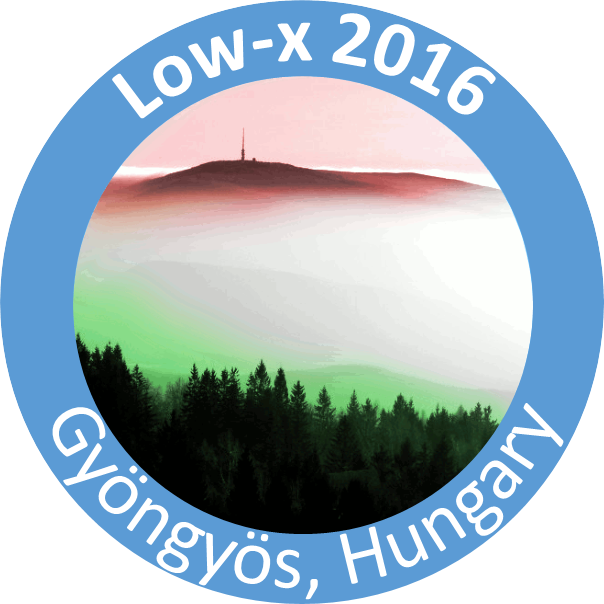}\\[1cm]
Search for a new baryonic state decaying to $pK^0_S$}
\author{P. J. Bussey\\
School of Physics and Astronomy, University of Glasgow,\\ 
Glasgow G12 8QQ, United Kingdom.\\[1ex]
\\ for the ZEUS Collaboration
}

\begin{document}

\fontfamily{lmss}\selectfont
\maketitle

\begin{abstract}
We report on a new ZEUS search for a narrow state decaying into
$p(\bar{p})K^0_S$, which was previously claimed by the ZEUS collaboration.
In the present search, which uses much increased statistics, no evidence for 
this state is found.  Limits on the cross section for such a state are given.
\end{abstract}

\section{Introduction}

\begin{figure}[t]
\begin{center}
~\\[-0.1\linewidth]
\includegraphics[width=0.45\textwidth]{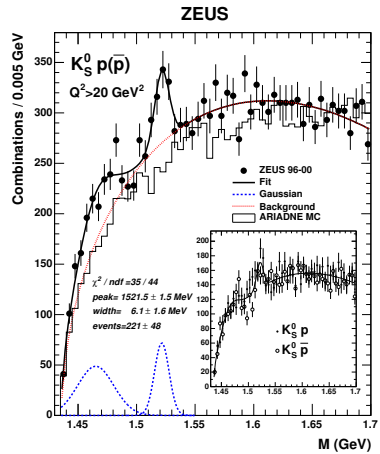}
\end{center}
\caption{The (anti)proton-kaon mass spectrum obtained by ZEUS in its
earlier publication, interpreted as showing a narrow state at
1.51215 GeV.
}
\label{fig0}
\end{figure}

During the early 2000 years, several experiments reported the
discovery of ``exotic'' hadronic particles that apparently consisted
of more than the normal two or three quarks or antiquarks.  The
results of the different experiments were not all consistent, and the
area was left unresolved.  Among the reported results was a paper from
the ZEUS collaboration~\cite{Z1}, in which the channel $pK_S^0$ was
studied together with the corresponding antiproton state.  Evidence
was given (fig.~\ref{fig0}) for a narrow peak in this channel, at a
mass of 1.5215 GeV, using data taken in the first running period of HERA
during the years 1996-2000.  Such a state would be a pentaquark state
$uudd\bar{s}$, and was consistent with the $\Theta$ state reported by
other groupss, although it was not confirmed by the H1 experiment at
HERA~\cite{H1}.

     Recently, the LHCb collaboration have announced the discovery of
two pentaquark states at the much higher masses of 4.38 and 4.45
GeV~\cite{LHCb}. The time was therefore ripe for a re-examination of
the state reported by ZEUS.  This presentation, whose results have
been published in more detail~\cite{Z2}, gives an analysis of the same
channel using the much larger data set collected during the second
HERA run period 2003-2007. In addition, the charged-particle tracking
system was upgraded and used improved analysis software.

\section{Analysis procedure}
A data set corresponding to 358 pb$^{-1}$ was used in the present
work, approximately three times the previously used integrated
luminosity.  The proton was identified using ionisation deposited in
the ZEUS central tracking detector (CTD) as before, augmented by
similar information available from the new silicon micro vertex
detector (MVD).  In the following account,any mention of a $pK^0_S$
state always includes the  consideration of the corresponding charge
conjugate state$\bar{p}K^0_S$.  Protons at 920 GeV were collided with
electrons or positrons at 27.5 GeV, and the state under investigation
was sought in deep inelastic scattering events with a well-identified
scattered electron or positron. The virtuality of the exchanged photon
was chosen to be in the range $20 < Q^2 < 100$ GeV$^2$, and the
scattered electron or positron had to have an minimum energy of 10
GeV.  The final-state particle had to be well contained within the
detector, as determined by the sum of their measured energy and
longitudinal momentum.

The $K^0_S$ contained in the sought state was identified in the normal way by
observing two charged tracks of opposite sign, produced in the region
of the detector with good tracking identification, and forming a
common vertex such that the reconstructed $K^0_S$ pointed towards the
primary vertex of the event.  The requirement was made for a decay
length of at least 0.5 cm, projected on to the plane transverse to the
beam line.  Each contributing track had to have a transverse momentum
of at least 0.15 GeV, and that of the combination had to be at least
0.3 GeV.  Cuts were made to eliminate converted electrons and lambda
baryons.  A clean $K_S^0$ signal was seen with negligible background,
correponding to 0.31 million events with a $K_S^0$

Proton identification was performed using the ionisation energy in the
CTD and the MVD, and was used to select proton candidates in the
momentum range 0.2 to 1.5 GeV. It was applied to all charged tracks
that passed through both detector system, ignoring tracks that had
been assigned to $K^0_S$ mesons. The two detectors were used to
confirm each other's ionisation signals and provide a satisfactory
proton identification when both detectors were employed together.  The
probability for identifying a given proton correctly varied from 0.8
at low $p_T$ values, falling to 0.2 at the upper end of the accepted
range.  Pion contamination was evaluated using a selected $K_S^0$
sample, and was 10-100 times better in the present measurement than in
the earlier measurement.  Pions identified as protons contribute to
the background in the mass spectrum under study.

\begin{figure}[t!]
\begin{center}
~\\[-0.31\linewidth]
\includegraphics[width=0.9\textwidth]{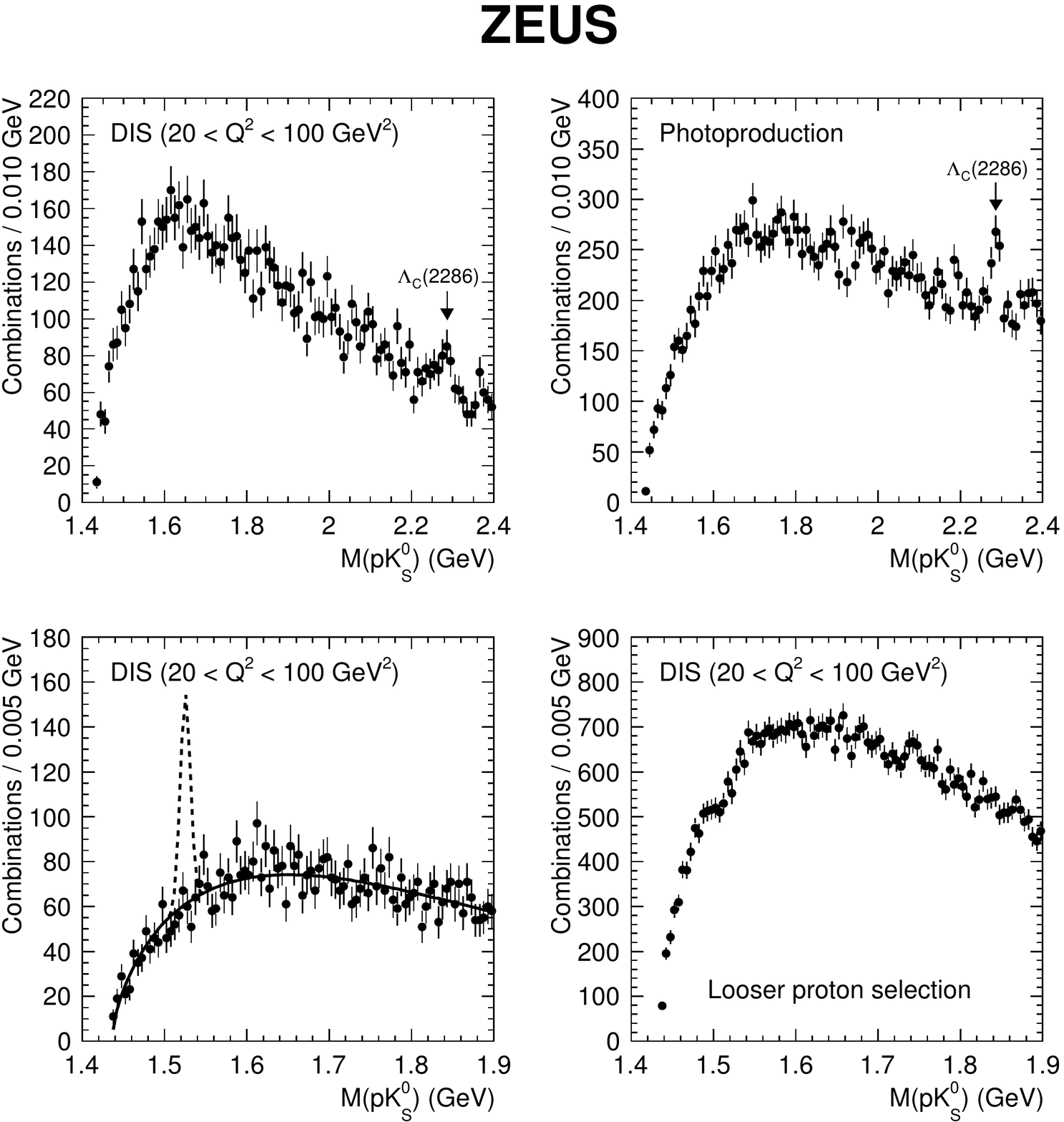}
\\[-0.535\linewidth]
(a)\hspace*{0.6\linewidth}(b)
\\[0.39\linewidth]
(c)\hspace*{0.6\linewidth}(d)
\\[0.1\linewidth]
\end{center}
\caption{(a) (Anti)proton-kaon mass spectrum measured in DIS events with
event selections as in the text; (b) mass spectrum measured in photoproduction events; (c) mass spectrum as (a) with expanded horizontal scale; (d) mass spectrum measured under conditions resembling those of  the earlier ZEUS result. } 
\label{figr}
\end{figure}

In order to verify that a narrow baryon-like signal was detectable
using the present method, the $pK_S^0$ mass spectrum was measured for a 
large sample of photoproduction events obtained with ZEUS.  A clear signal
for the $\Lambda_c$ baryon at 2.3 GeV was observed.

\section{Results}

Figure \ref{figr} presents $pK_S^0$ mass spectra obtained in the
present search.  Figure~\ref{figr}(a) shows the spectrum for a broad
range of mass values, and exhibits no narrow features except for an
indication of the $\Lambda_c$ baryon at 2.3 GeV, which is seen more
clearly in photoproduction events (b). Figure~\ref{figr}(c) shows the
same mass spectrum as (a), but on an expanded scale, and includes the
size of a signal at 1.5215 GeV that would be expected if the fitted
feature in the previous ZEUS analysis were found in the present data
sample. In (d), the shape of
the mass spectrum is presented under conditions of event selection
that resembled those of the previous ZEUS analysis.

It is evident that the previously claimed narrow state is not
confirmed in the present data.  Instead, it would seem that the shape
of the mass spectrum in the previous data was not modelled
appropriately, and  the apparent presence of a narrow peak was a
consequence of this together with a statistical fluctuation on the
data.

Given this conclusion, the ZEUS Collaboration have calculated
experimental limits on the production of a narrow state $X$ in the in
the range of mass values that have been measured.  Upper limits for
the cross section times the branching ratio into $pK_S^0$ are
evaluated for three hypotheses on the width of $X$: a fixed 
value of 6.3 MeV, corresponding to the width of the fitted state in
the earlier ZEUS paper, a mass-varying width corresponding to the
ZEUS detector resolution, and also twice this value.  The final case
roughly approximates to the measurement resolution of the H1 detector,
in order to make a comparison to the published H1 cross-section limits~\cite{H1}.

The hypothesised state $X$ was simulated using  the Monte Carlo
generator RAPGAP 3.1030, and the normal GEANT-based
simulation of the ZEUS apparatus.  The known state $\Sigma^+(1189)$,
which decays into $pK^0$, was replaced by a similar state whose mass
was assigned a series of values in data sets that covered
the relevant range of masses under study.  The normal branching ratios
for the kaon decay into $K_S^0$ and into $\pi^+\pi^-$ were taken, and
.  Cross section limits were calculated for production within the
kinematic limits $20 < Q^2 < 100$ GeV$^2$, $0.5 < p_T(X) < 3.0$ GeV/c,
$|\eta(X)| < 1.5$.  A polynomial-based background spectrum shape was
used.  Systematic effects gave an uncertainty of approximately 10\% on
the resulting upper limits, the biggest contribution coming from the
modelling of the transverse momentum of the state $X$.

\begin{figure}[t]
\begin{center}
~\\[-0.66\linewidth]
\includegraphics[width=0.9\textwidth]{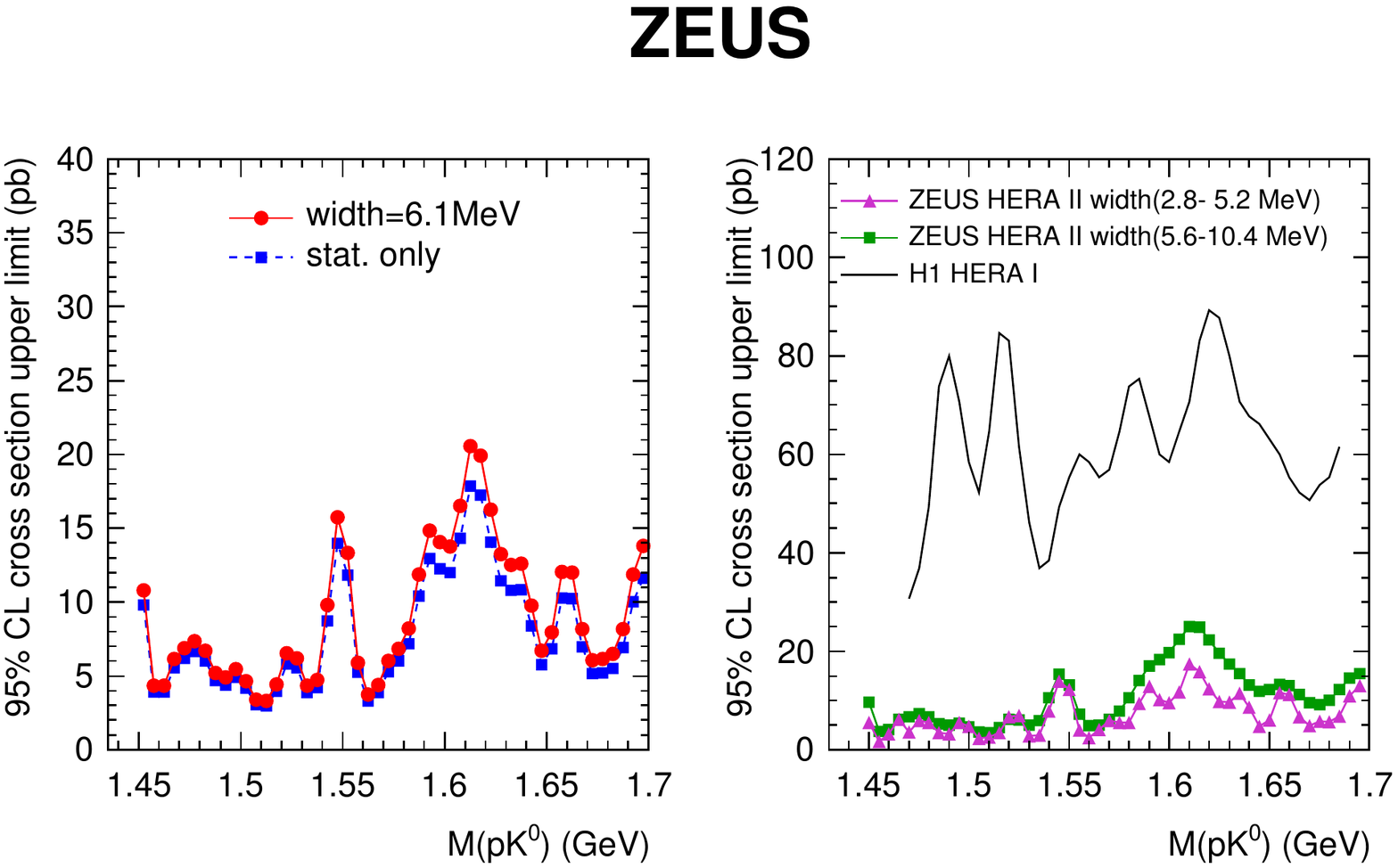}
\\[-0.30\linewidth]
(a)\hspace*{0.6\linewidth}(b)
\\[0.3\linewidth]
\end{center}
\caption{95\% confidence level upper limits for the production cross
section of a state $X$ times its  branching ratio for $X\to pK_S^0$
for (a) fixed 6.1 MeV width and (b) detector-resolution width models.}
\label{figs}
\end{figure}

The resulting cross-section limits at the 95\% confidence level are
shown in fig.~\ref{figs}, in comparison  with the higher
values that had been obtained by H1.

\section{Summary}
A new search has been made, by the ZEUS Collaboration at HERA, for a
narrow state decaying into $p(\bar{p})K^0_S$ whose observation was
previously claimed by the ZEUS collaboration.  In the present search,
which uses much increased statistics, no evidence for this state is
found.  Limits on the cross section for such a state are given.


\begin{thebibliography}{10}

\bibitem{Z1}
ZEUS Collaboration, Phys.\ Lett.\ {\bf B 591},  7 (2004).
\bibitem{H1}
H1 Collaboration, Phys.\ Lett.\ {bf B 639}, 202 (2006).
\bibitem{Z2}
ZEUS Collaboration, Phys.\ Lett.\ {\bf B 759},  446 (2016).
\bibitem{LHCb}
LHCb Collaboration, Phys.\ Rev.\ Lett.\ {\bf 115}, 072001 (2015).
\end{thebibliography}
\end{document}